\documentclass[review]{article}

\usepackage{lineno,hyperref}
\usepackage{amsmath}
\usepackage{graphicx,float}
\usepackage{amsmath,mathtools,booktabs}  
\usepackage{amssymb}  
\usepackage{xcolor,url}
\usepackage{pdflscape}
\usepackage{epstopdf} 
\usepackage{subfigure} 
\usepackage{amsmath,theorem}
\newtheorem{prop}{Proposition}
\newtheorem{cor}{Corollary}

\newtheorem{prob}{Problem}

\usepackage{natbib}
\setcitestyle{authoryear,open={(},close={)}}
\newcommand{\react}[1]{\ensuremath{\xrightharpoonup{\hbox{\makebox[.5cm][c]{\scriptsize $#1$}}}}}
\newcommand{\revreact}[2]{\ensuremath{  \xrightleftharpoons[\hbox{\makebox[.5cm][c]{\scriptsize$#2$}}]{\hbox{\makebox[.5cm][c]{\scriptsize$#1$}}}}}

  \newcommand*{\EF}[1]{\textcolor{black}{#1}}

\modulolinenumbers[5]

\usepackage[scaled]{helvet}

\title{A feedback SIR (fSIR) model highlights advantages and limitations of infection-dependent mitigation strategies}

\author{Elisa Franco, \\ Department of Mechanical and Aerospace Engineering, \\ Department of Bioengineering, \\  Molecular Biology Institute, \\
University of California, Los Angeles}

\begin{document}

\maketitle

\begin{abstract}
Transmission rates in epidemic outbreaks vary over time depending on the societal and government response to infections and mortality of the disease. Non-pharmacological mitigation strategies such as social distancing and the adoption of protective equipment aim precisely \EF{reducing transmission rates by reducing infectious contacts. To investigate the effects of mitigation strategies on the evolution of epidemics, nonlinear transmission rates that are influenced by the levels of infections, deaths or recoveries have been included in many variants of the classical SIR model.  This class of models is particularly relevant to the COVID-19 epidemic, in which the population behavior has been affected by the unprecedented abundance and rapid distribution of global infection and death data through online platforms. This manuscript revisits a SIR model in which the reduction of transmission rate is due to knowledge of infections. Through a mean field approach that assumes individuals behave like molecules in a well-mixed solution, one derives a time-varying reproduction number that depends on infection information through a negative feedback term that is equivalent to Holling type II functions in ecology and Michaelis-Menten functions in chemistry and molecular biology. A step-by-step derivation of the model is provided, together with an overview of methods for its qualitative analysis, showing that negative feedback structurally reduces the peak of infections. At the same time, feedback may substantially extend the duration of an epidemic. Computational simulations agree with the analytical predictions, and further suggest that infection peak reduction persists even in the presence of information delays. If the mitigation strategy is linearly proportional to infections, a single parameter is added to the SIR model, making it useful to illustrate the effects of infection-dependent social distancing.} 
\end{abstract}




\section{Introduction}

Compartment models are widely used to capture the long-term temporal evolution of epidemic outbreaks. Like mean-field models in physics and chemistry, compartment models assume a well-mixed population and capture average interactions patterns. The population is binned in distinct categories (the compartments), that at a minimum include those susceptible to disease (S), those who become  infected  (I), and those who recover (R), like in the well-known SIR model by Kermack and McKendrick~\citep{kermack1927contribution,hethcote2000mathematics}.  Because the SIR model is not suited to capture epidemics with a long incubation time, a large population of asymptomatic individuals, and high lethality, many SIR variants with additional compartments have been developed and tailored to model specific epidemic outbreaks~\citep{capasso1978generalization,bootsma2007effect,giordano2020modelling}. In the context of the COVID-19 pandemic, SIR-like models have been used to forecast local outbreaks~\citep{bertozzi2020challenges} and to formulate recommendations for suppression and mitigation strategies~\citep{kruse2020optimal,bin2020fast,casella2020can,della2020network}.

As the capacity of COVID-19 testing has increased, infection, recovery, and death data have become available at the local and global level with unprecedented speed thanks to online dashboards, apps, and media reports~\citep{Ensheng2020,ItalyCOVID19,Prasse2020}. This information has influenced the behavioral choices of the public, and has been essential for governments to make critical decisions in regards to suppression and mitigation policies. While these policies can successfully quench the epidemic with an \emph{open loop} approach that discounts recent data~\citep{bin2020fast,sadeghi2020universal}, strategies that respond in \emph{closed loop} to trends in the current level of infections, deaths, or recoveries are more likely to be accepted or even spontaneously adopted by the population. 

\EF{Compartment models have been adapted to capture the effects of societal responses (such as mitigation strategies, behavioral changes, and vaccinations) that are influenced in \emph{closed loop} by epidemic information. This has been done by introducing nonlinear transmission functions, also known as ``behavioral functions'', which are chosen empirically to be smooth threshold/saturation or polynomial functions~\citep{capasso1978generalization,anderson1978regulation,korobeinikov2006lyapunov}.  One of the first contributions in this area is the SIR model variant described by Capasso and Serio to describe the cholera epidemics in Bari in 1973, in which the transmission rate includes a general nonlinear function of infections~\citep{capasso1978generalization}. This  function is further specified to decrease and saturate as infections increase, like Michaelis-Menten rates in biochemistry and Holling type II functions in ecology. The general influence of nonlinear transmission rates on the equilibria and dynamics of similar models were examined in~\citep{liu1986influence,liu1987dynamical}, and more recently in~\citep{korobeinikov2005non,kyrychko2005global,li2017dynamic,chapwanya2012enzyme,kumar2020deterministic}. Bootsma and Ferguson adopted a SEIR model in which a nonlinear Michaelis-Menten term captures the effects of death awareness on social interactions during the 1918 influenza epidemic in the United States~\citep{bootsma2007effect}. Similarly, SIS, SIR, and SEIRS models have been modified to capture the reduction of transmission and contact rates achieved by infection awareness programs~\citep{greenhalgh2015awareness,samanta2014effect,yu2017effects}. These models have also examined how infection awareness can reduce the susceptible fraction of the population~\citep{kiss2010impact,funk2009spread}, in particular by increasing vaccination rates~\citep{buonomo2008global}.}

\EF{This manuscript revisits, examines, and provides some novel results on a class of modified SIR models originally described by~\citep{capasso1978generalization}, in which the infection-dependent transmission rate introduces a feedback loop. Rather than being adopted based on empirical observations, here the nonlinear transmission rate is derived step-by-step using the law of mass action. It is assumed that individuals behave like particles in a well-mixed solution, and their interactions are modeled through equivalent chemical reactions that can be converted to ordinary differential equations (ODEs) by applying the law of mass action. Mitigation strategies such as social distancing and use of protective equipment are modeled as reactions that reduce successful infections at a rate that depends on current infection levels. Through a quasi-steady state argument, one obtains a nonlinear transmission rate parameter that includes a specific ``mitigation function'' term and is comparable to Michaelis-Menten functions. This approach produces a SIR model similar to the one described in (Section 6,~\citep{capasso1978generalization}). Because the nonlinear transmission function decreases as a function of infections, a \emph{negative} feedback loop emerges and it is convenient to adopt the nomenclature feedback SIR (fSIR).}

\EF{Most nonlinear epidemic models are positive and structurally bounded; stability analysis of the equilibria can be done via local (linearization) or global methods (typically Lyapunov functions), obtaining conditions for convergence to a disease-free equilibrium, or conditions for the occurrence of bifurcations in the presence of an endemic equilibrium (in which a fraction of the population remains infected)~\citep{liu1987dynamical,korobeinikov2005non,kyrychko2005global,greenhalgh2015awareness,weitz2020modeling}. Here the equilibria and the solutions of the fSIR model are examined (through established approaches) with focus on comparing the outcome in the presence and in the absence of mitigation. Specifically, I show that a broad class of infection-dependent mitigation functions makes it possible to reduce the peak of infections for any mitigation intensity. If the mitigation function depends linearly on infections, it is shown that the peak is also postponed for all positive mitigation parameters. These benefits of mitigation are however counterbalanced by the fact that the duration of the epidemic, measured as the time for which infections persist, may significantly increase -- an effect that is demonstrated with a simple linear approximation.  Computational simulations support the analysis reported, and indicate that  mitigation of the peak persists even in the presence of delay in the transmission of infection information, which induces a moderate retardation of the time at which infections peak. Finally, for purely illustrative purposes, I highlight that the fSIR model can qualitatively capture infection data of the COVID-19 pandemic for countries like the United Kingdom, the United States and Sweden, that opted for mitigation rather than suppression.}

\EF{This brief study of the fSIR model shows that it is a helpful tool to illustrate the effects of mitigation strategies in epidemics, with particular relevance to the COVID-19 epidemic that is characterized by rapid spread of information and fluctuations in social distancing patterns. While more complex models and data-driven parameter estimation are clearly needed for epidemic prediction~\citep{anastassopoulou2020data,giordano2020modelling,calafiore2020time}, simple yet rigorous models like the fSIR are valuable as they provide qualitative insights.  Only one term, the mitigation function, is needed in addition to the reproduction number to describe the evolution of the epidemic in the presence of infection-based mitigation strategies. This term reduces to a single parameter in the special case in which the mitigation strategy is a linear function of infections. Further, this model supports mitigation guidelines as it clearly shows that the infection curve can be flattened without postponing the peak, a misleading (and demotivating) scenario suggested by similar models that use a constant transmission rate. At the same time, the model highlights that policies relying exclusively on infection data to regulate social distancing can majorly extend the time required to reach a disease-free equilibrium.}

\subsection{Background: qualitative analysis of the non-dimensional SIR model}
\EF{The well-known SIR model is reviewed in this section to establish notation and background concepts~\citep{hethcote2000mathematics}.} It is assumed that the total population remains constant (birth and death processes are neglected) and the dynamics are driven by two key parameters: \\
\noindent 1) the disease transmission coefficient $\beta$, which depends on the social interactions among individuals (average daily contacts) and on the infection characteristics; the transmission rate is generally thought as the product of the average frequency of contacts between infected and susceptible and the likelihood that infection occurs given a contact;  \\
\noindent 2) the recovery coefficient $\gamma$, which captures the average time for recovery (or death) of infected individuals. The inverse $1/\gamma$ is also known as duration of infectiousness. Assuming the total population is $N$, the original SIR model is:
\begin{align}\label{SIR1}
\frac{dS}{dt}&= -  \frac{\beta}{N} I S,\\ \label{SIR2}
\frac{dI}{dt}&=\frac{\beta}{N} I S - \gamma I,\\ \label{SIR3}
\frac{dR}{dt}&= \gamma I.
\end{align}
\EF{For simplicity, here we do not model the possibility of reinfection of recovered individuals~\citep{hethcote1976qualitative}. The transmission coefficient is normalized by the total population size, because the number of new infections per unit time occur based on the average infectious contacts of each susceptible individual and does not depend on the total population size (this is also called standard incidence~\citep{hethcote2000mathematics,hethcote1976qualitative}).} \EF{Because the total population is assumed to remain constant, at any point in time $R=N-I-S$ and the model can be reduced to two ordinary differential equations (ODEs).} Further, the variables can be normalized by the total population  setting $s=S/N$, $i=I/N $ (and $r=R/N$); by rescaling  time  as $\tau=t\gamma$, the SIR model becomes non-dimensional, with a single coefficient $\mathcal{R}_0 = \beta/\gamma$, the well known reproduction ratio or \EF{reproduction number~\citep{hethcote1976qualitative}.} 
\begin{align}\label{SIRn1}
\frac{ds}{d\tau}&= - \mathcal{R}_0 i s, \\ \label{SIRn2}
\frac{di}{d\tau}&= (\mathcal{R}_0 s - 1) i.
\end{align}
\EF{Given initial conditions $s_0=s(0)$ and $i_0=i(0)$, the solutions $s(\tau)$ and $i(\tau)$ will be generally denoted as $s$ and $i$ with the assumption that these symbols indicate functions of time (unless otherwise noted).}
It is well-known that the solutions are non-negative and satisfy the conservation law $s+i+r=1$~\citep{hethcote1976qualitative}. Exact expressions for the solution have been computed~\citep{Harko14}. If there are no infected individuals ($i_0=0$), the system remains in the \EF{equilibrium $E_0=(s_0,0,r_0)$} because all derivatives are identically zero.  For any initial value of infections $i_0>0$, the solutions $s$ and $i$ are bounded and evolve in the invariant set \EF{$\mathcal{P}=\{0\leq s\leq s_0,\,0\leq i\leq 1, \,0\leq r\leq 1\}$}. This follows from the fact that $ds/d\tau\leq 0$, so $s(\tau)\leq s_0$, $\forall \tau\geq \tau_0$. The solutions and the admissible equilibria depend on the value of $\mathcal{R}_0$ and on the initial value of the susceptible population $s_0$.   

\EF{Case 1: $\mathcal{R}_0 s_0=0$. This occurs when the transmission rate or the initial susceptible population are equal to zero. In either case, $s(t)$ remains identically zero; $di/d\tau\leq 0$ becomes a linear asymptotically stable equation with zero as the only equilibrium.}

\EF{Case 2: $\mathcal{R}_0 s_0<1$. In this case} the infected population is non-increasing because $di/d\tau\leq 0$, thus the epidemic does not start (the system reaches \EF{an equilibrium $\tilde{E}=(\tilde s,0,\tilde r)$}). 

\EF{Case 3: $\mathcal{R}_0 s_0<1$. In this case} $di/d\tau$ initially increases, reaches a peak when $s=s_{crit}=1/\mathcal{R}_0\leq s_0$, and finally decreases to zero. \EF{The equilibrium in this case is $E=(\bar s, 0, \bar r)$.} Because $s_0\leq 1$, $\mathcal{R}_0 s_0>1\Rightarrow\mathcal{R}_0 >1$. 
For any positive $i_0$ and $\mathcal{R}_0 s_0>1$, the relation between susceptible and infected can be computed exactly from the \EF{ratio of $di/d\tau$ and $ds/d\tau$~\citep{hethcote1976qualitative}:}
\begin{align*}
&\frac{di}{ds}=\frac{\mathcal{R}_0 s -1}{-\mathcal{R}_0 s} = -1 +\frac{1}{\mathcal{R}_0 s} \, \Rightarrow \, di= -ds + \frac{ds}{\mathcal{R}_0s}.
\end{align*}
\EF{Integrating we obtain the relation between $i$ and $s$:
\begin{align}\label{PhaseSpace} 
i&=i_0 + s_0 - s- \frac{1}{\mathcal{R}_0}\log\frac{s_0}{s}. 
\end{align}}
The peak of infections occurs when $s=s^*={1}/{\mathcal{R}_0}$ ($s=s^*$ yields $di/d\tau =0$). Substituting $s^*$ we find:
\begin{align}\label{Peak}  
i_{max}=i_0 + s_0 - \frac{1}{\mathcal{R}_0}\left(1+\log (s_0\mathcal{R}_0) \right),
\end{align}

with $\log (s_0\mathcal{R}_0) >0 $ because $s_0\mathcal{R}_0>1$ \EF{(we assume that prior to the start of the epidemic the recovered population is zero, thus $i_0+s_0=1$)}. From expression~\eqref{PhaseSpace}, by setting $\bar i=0$, we can also derive an implicit equation to find the equilibrium value of the \EF{susceptible} population:
\begin{equation}\label{SIRss}
 \log \frac{s_0}{\bar s} = \mathcal{R}_0 (1-\bar s), 
 \end{equation} 
which has one positive root (because $\bar s < s_0 \leq 1$ and $\mathcal{R}_0 >1$). In other words, the equilibrium susceptible population is positive (not all the population has become infected), unless $\mathcal{R}_0$ is unrealistically large.

\begin{figure}[htbp]
\centering
\includegraphics[width=\columnwidth]{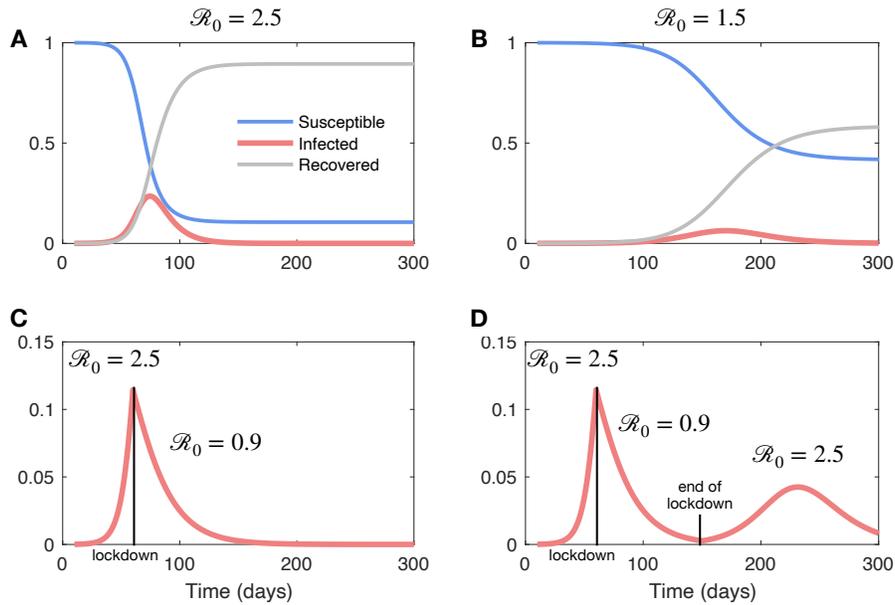}
\caption{A and B: Illustrative computational simulations showing the SIR dynamics for different values of (constant) reproduction coefficient $\mathcal{R}_0$. The plot in B illustrates how a lower value of $\mathcal{R}_0$ ``flattens the curve'' while also significantly delaying the infection peak. This illustration may be misleading to the public, because the introduction of suppression or mitigation measures causes the transmission rate constant $\mathcal{R}_0$ to vary in time. C: A lockdown scenario in which $\mathcal{R}_0$ switches from 2.5 to 0.9 after day 60. D: Lifting the lockdown at day $t_{\text{end}}=150$ causes the infections to increase again ($s(t_{\text{end}})\mathcal{R}_0>1$).}
\label{SIRBeta}
\end{figure}

\subsection{Flattening and reshaping the infection curve through suppression and mitigation policies}

\EF{The SIR model has been often used during the COVID-19 pandemic} to illustrate how a low reproduction number $\mathcal{R}_0$ (or a low transmission rate $\beta$)  has the effect of ``flattening the (infection) curve'', \emph{i.e.} reducing the infection peak while extending the duration of the epidemic. The simulations in Fig.~\ref{SIRBeta} A and B compare the SIR solutions for values of $\mathcal{R}_0=2.5$, which is close to recent estimates for the COVID-19 outbreak~\citep{earlyR0}, and $\mathcal{R}_0=1.5$. The infection peak is clearly reduced when $\mathcal{R}_0=1.5$, however the infection peak is also significantly delayed. The reproduction number depends on many factors, including societal habits and pharmacological interventions. In 2020, reducing the reproduction number of COVID-19 is only possible by controlling societal interactions, given the lack of approved vaccines and standardized medical treatment protocols~\citep{stewart2020control}.  

Suppression (lockdown) or mitigation (social distancing and adoption of Personal Protective Equipment, PPE) policies aiming to control and extinguish the epidemic may fluctuate over time to minimize their impact on society, thereby introducing fluctuations of $\mathcal{R}_0$~\citep{stewart2020control}. While useful to illustrate the concept and the effects of the reproduction number, Fig.s~\ref{SIRBeta} A and B do not represent temporal changes of $\mathcal{R}_0$ and are thus misleading to the public and to policymakers. During the COVID-19 epidemic, enormous research efforts are dedicated to a continuous estimation and forecasting of the reproduction number as a function of societal response~\citep{giordano2020modelling,bertozzi2020challenges,anastassopoulou2020data,kissler2020projecting}. 

As of mid 2020, the most successful strategy to manage COVID-19 \EF{was} full suppression of social interactions (lockdown); states such as China, South Korea, Italy, Spain, and France went on strict lockdown for more than two months, containing infections by Summer 2020. Qualitatively, the effects of a lockdown can be captured by a SIR model in which $\mathcal{R}_0$ rapidly changes from a high to a low value; the simulation in Fig.~\ref{SIRBeta}C illustrates the profile of infections under an abrupt change of $\mathcal{R}_0$ from 2.5 to 0.9 after 60 days from the start of the epidemic; the disease-free equilibrium is reached within a few months from the start of the suppression. However, ending lockdown measures too early can cause the epidemic to restart if $s(t_{\text{end}})\mathcal{R}_0>1$, as illustrated in  Fig.~\ref{SIRBeta}D, where the lockdown is completely lifted after 90 days~\citep{bertozzi2020challenges}. The success of lockdown is also tied to the ability to coordinate regulations and enforcement, and to sustain its major impact on the economy and on the mental health of the population. Due to the significant upfront ``cost'',  lockdowns \EF{are unpopular and difficult to enforce}.

\EF{Mitigation strategies have been adopted in many countries during the COVID-19 pandemic as a complement or replacement to lockdowns, and are thus an important phenomenon that should be included in mathematical models.} Mitigation means the reduction of large-scale public events, closure of certain businesses, and safe-at-home orders that could be classified as social distancing; mitigation efforts include the use of PPE such as masks, face shields, and gloves. Mitigation policies may become more restrictive or relax over time, depending on fluctuations of the contagion data, and on social and political climate. Restrictions to social interactions are likely to be more effective if they are tied to the reported infections or deaths, which increase the perceived risk of infection.  With fast spread of information about testing results~\citep{Ensheng2020,ItalyCOVID19,Prasse2020}, knowledge of \emph{infections} may be more helpful than deaths in quickly containing epidemics; because the average time to death for COVID-19 patients, for example, is 17 days~\citep{zhou2020clinical}, reliable lethality information may only be available with a significant delay. 

\EF{The rest of this manuscript derives and revisits an SIR model that qualitatively captures mitigation strategies and societal responses based on knowledge of infections, which introduce feedback in the epidemic process.}

\section{Results}

\subsection{Mitigation policies yield nonlinear transmission rate parameters} 

\EF{Here I provide a simple step-by-step derivation of a SIR model in which the transmission rate parameter varies as a function of infection-based mitigation policies, reproducing the empirical model described by~\citep{capasso1978generalization}. It is assumed that individuals behave like molecules in a well-mixed solution and interact through equivalent chemical reactions. The corresponding ODEs are derived using the law of mass action in chemistry. A related mean-field approach, that considers individuals as agents that interact with a limited foraging radius has been considered in~\citep{kolokolnikov2020law}, obtaining an exponential saturating transmission rate. In the context of predator-prey models, in~\citep{dawes2013derivation} the population-level Holling's functional responses is derived in a limit scenario starting from individual-level stochastic interactions.}

First, a contagion may occur when a susceptible individual ($S$) and an infected individual ($I$) are in spatial proximity for some time (associated or contact state $C$); this encounter may then result in two infected individuals. This can be modeled using the equivalent chemical reactions:
\begin{align*}
S+I\revreact{\rho^+}{\rho^{-}} C \react{\phi} 2 I,
\end{align*}
 where $\rho^+$ and $\rho^-$ are the rates of association and dissociation of a susceptible and an infected individual, and we can associate \EF{$\phi$ with the daily rate at which individuals that have been exposed become infected. The law of mass action converts reactions like the one above to ODEs in which variables are concentrations of reactants and products, computed by dividing the number of molecules by the reaction volume. Similarly, here one can derive an ODE for the fraction of individuals in each compartment by dividing the number of individuals by the total population. The ODE describing the kinetics of the fraction of individuals ($c$) in the associated state ($C$) is:}
\begin{align*}
\frac{dc}{dt} & = \rho^+ s \cdot i - (\rho^- +\phi) c. 
\end{align*}
Because contacts occur on an hourly or daily basis, which is much faster than timescale of the epidemic, it is sensible to assume $dc/dt=0$ and derive an expression for the equilibrium level of associated individuals:
\[ \bar c = \frac{\rho^+}{\rho^- +\phi} s \cdot i.\]
This value of $\bar c$ is intended to represent a dynamic equilibrium at the population level, so it indicates the average number of contacts per day.
With this definition, the transmission rate $\beta$ introduced in model~\eqref{SIR1}-\eqref{SIR2} is:
\[ \beta=\frac{\rho^+ \phi }{\rho^-+\phi},\] 
where $\phi$ is the probability of infection per contact, and ${\rho^+}/(\rho^-+\phi)$ is the average number of contacts per day  \EF{per individual}, a definition that is consistent with the \EF{literature~\citep{hethcote2000mathematics}}. \footnote{This definition of $\beta$ can be verified by using the law of mass action to write the ODEs of $s$ and $i$. For example \[ \frac{ds}{dt}=-\rho^+ s\cdot i - \rho^- c,\] in which  $c$ has to be  replaced by its equilibrium value $\bar c$.} The corresponding (non-dimensional) reproduction coefficient can be computed as earlier $\mathcal{R}_0=\beta/\gamma$. \EF{Note that if $\rho=0$ and $\phi$ is slow, with this approach we would recover the SEIR model~\citep{hethcote2000mathematics}, where the ``contact'' species $C$ corresponds to the exposed category $E$. Here we will assume that the parameter $\phi$ is large enough that the contact $\bar c$ can be neglected in the overall mass balance; if this were not the case, then $\bar c$ must be explicitly included in the mass equation $s+i+\bar c + r =1$.}

In the presence of mitigation policies that discourage association of individuals, \emph{i.e.} social distancing, the level of individuals in associated state $C$ should decrease. This can be modeled by additional, fast dissociation process that depends on the known infection levels through a rate parameter $\psi(I)$:
\[ C \react{\psi(I)}  S+I. \] 
For this to be a well-posed reaction, we require the distancing parameter $\psi(I)$ to be a non-negative, non-decreasing function of $I$, with $\psi(0)=0$.  
With this model for dissociation, individuals in state $c$ evolve according to the ODE:
\begin{align*}
\frac{dc}{dt} & = \rho^+ s \cdot i - (\rho^- +\phi) c - \psi(i) \cdot c, 
\end{align*}
which equilibrates to:
\[ \bar c = \left(\frac{\rho^+}{\rho^- +\phi}\right) \frac{1}{1+\kappa(i)}s \cdot i, \quad \kappa(i) = \frac{\psi(i)}{\rho^-+\phi}. \]
With this equilibrium value for the average contacts, we derive a time-varying expression for the reproduction number that depends on the infection levels: 
\begin{align} \label{R0time}
\mathcal{R}(i)=\mathcal{R}_0\frac{1}{1+\kappa(i) }. 
\end{align}
The {function} $\kappa(i)$ is in units of /time/individual (or fraction of individuals, the equivalent of a normalized concentration in chemical reaction networks). Thus $\mathcal{R}(i)$ is non-dimensional like $\mathcal{R}_0$.

Expression~\eqref{R0time} is equivalent to Holling type II functions in ecology, and Michaelis-Menten/Hill functions  in chemical kinetics, and indicates that under a policy in which social distancing depends on the infection levels, the reproduction number $\mathcal{R}(i)$ decreases as the infection numbers raise~\citep{capasso1978generalization,bootsma2007effect,li2017dynamic}.  One can think about the feedback term $1/(1+\kappa(i))$ as a reduction of the duration or frequency of infectious contacts introduced by social distancing policies. 

Another successful approach to mitigate the spread of contagions is to recommend the use of PPE such as masks and gloves when the number of infected individuals increases. A simple way to model the average effect of PPE is to assume a change in the likelihood of infection following a contact:  
\begin{align*}
S+I\revreact{\rho^+}{\rho^{-}} C \react{\phi(I)} 2 I,
\end{align*}
with $\phi(I)$ being a decreasing function of the level of infections: the more contagions are known, the more widespread is the use of PPE, the lower the chance of becoming infected. One ought to assume that in the absence of information on infections, the natural infection probability is recovered, \emph{i.e.} $\phi(0)=\phi$. A suitable function is:
\[\phi(I)=\frac{\phi}{1+\xi(I)}, \]
 with $\xi(0)=0$, and $\xi(I)$ non-negative, non-decreasing. With a timescale separation argument one can find the average daily level of (normalized) infectious contacts:
 \[ \bar c = \frac{\rho^+}{\rho^-+\xi(i)}. \]
 With this expression, the time-varying reproduction number is: 
 \[ \mathcal{R}(i)=\mathcal{R}_0\frac{1}{1+\kappa(i)},\quad \kappa(i)=\frac{\rho^- \xi(i)}{\rho^-+\phi}.\]
 This result is identical to equation~\eqref{R0time} if we take  $\xi(i)=\psi(i)/\rho^-$. 
 For this reason, from now on we will use the time-varying reproduction number~\eqref{R0time} as a general expression to model the effects of infection-aware mitigation on the dynamics of an epidemic. In the rest of the manuscript,  $\kappa(i)$ will be called \emph{mitigation function}.


\subsection{The feedback SIR model}
   
With infection-aware mitigation policies, the non-dimensional SIR model~\eqref{SIRn1}-\eqref{SIRn2} becomes the feedback SIR (fSIR) model:
\begin{align}\label{fSIRn1}
\frac{ds}{d\tau}&= -\mathcal{R}_0\frac{1}{1+\kappa(i) } s i = - \mathcal{R}(i) s i\\ \label{fSIRn2}
\frac{di}{d\tau}&=\left(\mathcal{R}_0\frac{1}{1+\kappa(i) } s -1\right)i=(\mathcal{R}(i) s -1) i. 
\end{align}

\begin{figure}[htbp]
\centering
\includegraphics[width=8cm]{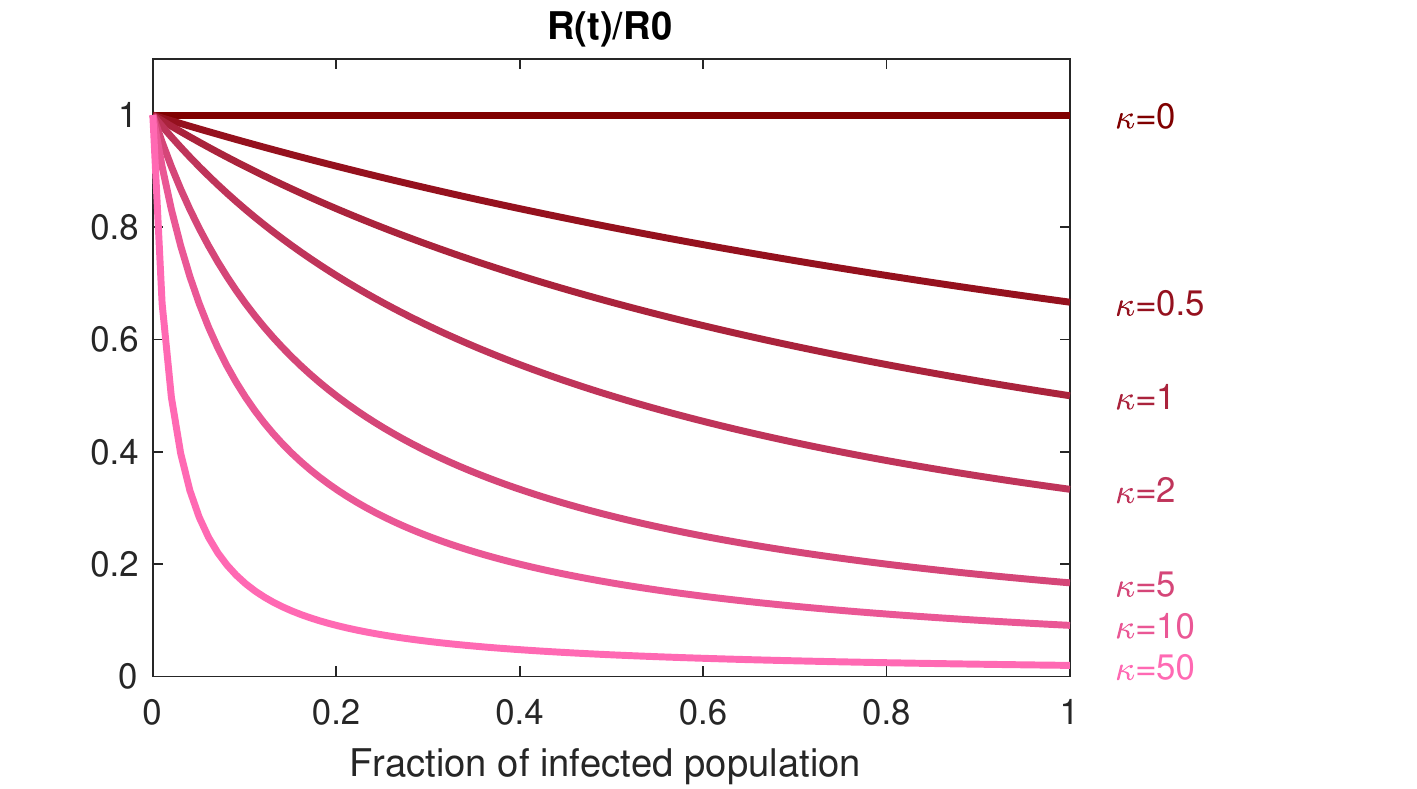}
\caption{In the case of mitigation function linearly proportional to infections, $\kappa(i)=\kappa\cdot i$, the infection-dependent reproduction number~\eqref{R0time} is monotonically decreasing as a function of infections, for any choice of  $\kappa\geq 0$.}
\label{Lambda}
\end{figure}

\EF{In the fSIR model the transmission rate is the nonlinear function $\mathcal{R}(i)=\mathcal{R}_0\frac{1}{1+\kappa(i) }$; we assume the mitigation function $\kappa(i)$ is a non-negative, non decreasing function of $i$, with $\kappa(0)=0$.} For the simple case in which $\kappa(i)=\kappa i$ (\EF{mitigation function} linearly proportional to infections),  $\mathcal{R}(i)$ decreases monotonically as a function of $i$, and it decreases more steeply for large values of $\kappa$, as illustrated in Fig.~\ref{Lambda}. The larger $\kappa$, the smaller the value of $i$ that induces a significant reduction in $\mathcal{R}_0$ (i.e.  distancing and PPE are adopted in response to a very small outbreak). For example, a value of $\kappa=2$ results in $\mathcal{R}(i)=\mathcal{R}_0/2$ when $i=0.5$; a value of $\kappa=10$ cuts in half $\mathcal{R}_0$ much sooner, when $i=0.1$.

The mitigation function $\kappa(i)$ models the average population response to knowledge of current infection numbers, in relation to typical interaction patterns; this coefficient could also be used to model the collective ``trust'' in infection information.  \EF{For $\kappa(i)=0$, i.e. there is no reaction/policy, nor trust on infection data, then $\mathcal{R}(i)=\mathcal{R}_0$. (Similarly, if there are no infections and $i=0$, then we have no change in $\mathcal{R}(i)=\mathcal{R}_0$ because $\kappa(0)=0$).}  

The time varying reproduction number $\mathcal{R}(i)$ introduces a \emph{negative} feedback loop in the epidemic model, because captures the fact that society mitigates interactions in response to an increase of infections, thereby reducing the reproduction number. \EF{This expression for $\mathcal{R}(i)$ also models the return to typical interaction patterns when infections are no longer present.}

\subsection{\EF{Properties} of the fSIR model}
\subsubsection{Analysis of equilibria} 

Local equilibrium analysis and and global stability analysis of SIR models with nonlinear transmission rates has been extensively carried out in the literature~\citep{capasso1978generalization,liu1987dynamical,korobeinikov2005non}. \EF{A brief discussion of the local stability of equilibria is reported below for illustrative purposes.} 

If $i_0=0$ ($r_0=0$), the system remain at the \EF{equilibrium $E_0=(s_0,0,0)$} because all derivatives are identically zero.  For any $0<i_0<1$, the solutions are bounded and evolve in the invariant set \EF{$\mathcal{P}=\{0\leq s\leq s_0,\,0\leq i\leq 1,\,0\leq r\leq 1\}$}.   If $\mathcal{R}_0 s_0\leq1+\kappa(i_0)$, the infected population is non-increasing because $di/d\tau\leq 0$, the epidemic does not start and the system reaches an equilibrium $\bar{E}=(\bar s,0,\bar r)$. Like in the SIR model, because $s_0\leq 1$, for the epidemic to start it is necessary that $\mathcal{R}_0 >1+\kappa(i_0)$. 

\EF{If $\mathcal{R}_0 s_0>1+\kappa(i_0)$,  $di/d\tau>0$ until the susceptible population decreases to the value $s=s_{crit}=(1+\kappa(i_{max}))/\mathcal{R}_0>\mathcal{R}_0$ at which $i(\tau)=i_{max}$. As the susceptible population continues to decrease, so does the infected population and the system reaches an  equilibrium $E=(\bar s, 0, \bar r)$. 
\begin{prop} Assume $\mathcal{R}_0 s_0>1+\kappa(i_0)$. Any equilibrium $\bar E=(\bar s, 0, \bar r)$ is \EF{locally stable}. \end{prop}
\emph{Proof}  The Jacobian of the fSIR model is:
\begin{align} 
J& =\begin{bmatrix} - {\bar i}\mathcal{R}(\bar i) & - \bar s \frac{d}{di}\mathcal{R}(\bar i)\\
{\bar i}\mathcal{R}(\bar i) & \bar s \frac{d}{di}\mathcal{R}(\bar i)-1
\end{bmatrix}
=\mathcal{R}_0\begin{bmatrix} - \frac{\bar i}{1+\kappa(\bar i)}  & - \frac{\bar s(1+k(\bar i)-\bar i \frac{d\kappa(\bar i)}{di})}{(1+\kappa(\bar i)^2)}\\
\frac{\bar i}{1+\kappa(\bar i)} & \frac{\bar s(1+k(\bar i)-\bar i \frac{d\kappa(\bar i)}{di})}{(1+\kappa(\bar i))^2}-\frac{1}{\mathcal{R}_0}
\end{bmatrix}.
\end{align} 
At the equilibrium $E=(\bar s, 0, \bar r)$, because $\kappa(0)=0$ by assumption, $J$ is identical to the Jacobian of the SIR model:
\[ J_0=\begin{bmatrix}  0  & - \bar s\\
0 & \mathcal{R}_0 \bar s-1 
\end{bmatrix},\] 
which is a \EF{stable matrix for any value of $\mathcal{R}_0\geq 0$ as long as  $\mathcal{R}_0 \bar s<1$ (at equilibrium it must be true that $\mathcal{R}_0 \bar s<1$)}.  $\hfill \square$}

\EF{Note that by assuming different nonlinear transmission rates, and including birth, death, and reinfection rates, \emph{endemic} equilibria may emerge in which the equilibrium infectious population is positive, and bifurcations may occur~\citep{liu1987dynamical,hethcote2000mathematics,korobeinikov2005non}.}

\subsubsection{\EF{Analysis of the solutions: advantages and disadvantages of infection-based feedback}}

\EF{By assuming that the derivative of the nonlinear transmission rate $\mathcal{R}(i)$ is bounded and has a maximum at $i=0$, \citep{capasso1978generalization} demonstrate global positivity, uniqueness, and global stability of the solutions for the fSIR model; these results can be extended to similar models that include birth, death, and reinfection rates, and assumptions on the transmission rate can be relaxed as reviewed in~\citep{korobeinikov2006lyapunov}.  Here $\kappa(i)$ is assumed to be non-decreasing, and zero for $i=0$, yielding a nonlinear transmission rate that is non-increasing and equal to $\mathcal{R}_0$ for $i=0$~\citep{korobeinikov2006lyapunov}. In this case, it is shown that the peak of infections is always reduced in the presence of distancing. I will also summarize results that exist for the case in which the mitigation function is linear ($\kappa(i)=\kappa \cdot i$), and provide some additional qualitative result in regards to the time at which the infection peak occurs.}

\begin{prob}\label{fSIRProblem}
The fSIR model~\eqref{fSIRn1}-\eqref{fSIRn2} with initial conditions $s_0\geq 0$, $i_0>0$, $r_0\geq0$, and $s_0\mathcal{R}_0>1+\kappa(i_0)$ defines an initial value problem (IVP) with non-negative solutions. We assume the mitigation function $\kappa(i)$ is a non-negative, non-decreasing function with $\kappa(0)=0$, and we look for properties of the solutions of this IVP that hold for any $\mathcal{R}_0$. These properties will be contrasted to the limit case $\kappa(i)=0$ that corresponds to the IVP defined by the SIR model~\eqref{SIRn1}-\eqref{SIRn2}. \\ The solution for $\kappa(i)=0$ as well as its features will be denoted with the superscript $^0$ (\emph{i.e.} if $\kappa(i)=0$, $i^0(\tau)=i(\tau)$).
\end{prob}

\EF{{\bf Nonlinear mitigation function:}}
In the general case of a nonlinear mitigation function $\kappa(i)$, I will show that the peak of infections in the fSIR model is smaller than the infection peak for the SIR model, for any non-negative $\kappa(i)$; to the best of my knowledge, this is a novel result. No assumption is needed on the boundedness of the derivative of $\kappa(i)$ like in~\citep{capasso1978generalization}.
 
\begin{prop}\label{PeakProp}
In Problem~\ref{fSIRProblem}, for any $\mathcal{R}_0$ and for any $\kappa>0$, we have: 
\[   i_{max} < i^0_{max}.\]  
\end{prop}
\emph{Proof} 
Following the same approach used to derive~\eqref{Peak}, the peak of infection for the fSIR model can be \EF{estimated as follows:}
\[\frac{di}{ds} =-1 + \frac{1+\kappa(i)}{\mathcal{R}_0s}.\] 
We then obtain the infinitesimal expression: 
\begin{align}\label{fSIRPeakInfinitesimal}
di& =  -ds + \frac{ds }{\mathcal{R}_0s} +   \kappa(i) \,\frac{ ds }{\mathcal{R}_0s}, 
\end{align}
\noindent in which the last term cannot be easily integrated, but it can be replaced by a simpler expression. \EF{Rearranging the terms of the ODE \eqref{fSIRn1} we find:} 
\[ \frac{1}{\mathcal{R}_0} \frac{ds}{s} = - \frac{i}{ 1 + \kappa(i)}d\tau,\]
which can be substituted in  the last term of  equation~\eqref{fSIRPeakInfinitesimal}:
\[ di =  -ds + \frac{ds }{\mathcal{R}_0s}  -  i \frac{\kappa(i) }{ 1 + \kappa(i)}d\tau,\]
thus we obtain the expression:
\begin{align}\label{fSIRPeakNew}
\EF{i}& =  s_0+i_0 - s + \frac{1}{\mathcal{R}_0}\log \frac{s}{s_0}  - \int_0^\tau i \frac{\kappa(i)}{ 1 + \kappa(i)}d\sigma, 
\end{align}
The infection peak occurs at $s_{crit}=(1+\kappa(i_{max}))/\mathcal{R}_0$, which can be substituted in equation~\eqref{fSIRPeakNew}:
 \begin{align}\label{fSIRPeak}
i_{max}&(\tau_{max}) =  s_0+i_0 - \frac{1+\kappa(i_{max})}{\mathcal{R}_0} +\\ \nonumber
&  + \frac{1}{\mathcal{R}_0}\log \left( \frac{1+\kappa(i_{max})}{\mathcal{R}_0s_0}\right)  - \int_0^{\tau_{max}} i \frac{\kappa(i)}{ 1 + \kappa(i)}d\sigma. 
\end{align}
When $\kappa(i)=0$ we recover the original SIR infection peak expression~\eqref{Peak},  here denoted as $i^0_{max}$. The difference between the peak value~\eqref{fSIRPeak} and $i^0_{max}$ (the peak when $\kappa(i)=0$) is:
\begin{align*} 
& i_{max} - i^0_{max}  = -\frac{1}{\mathcal{R}_0}\left(\kappa(i_{max}) - \log(1+\kappa(i_{max})) \right) - \\
 & \qquad \qquad- \int_0^{\tau_{max}}  i \frac{\kappa(i)}{ 1 + \kappa(i)}d\sigma.
\end{align*}
Because $\log(1+x) < x$ for any $x>0$, and because the last integral is strictly positive, we conclude that $i_{max} < i^0_{max}$ for any $\kappa>0$. $\hfill \square$

\begin{cor} 
\EF{In Problem~\ref{fSIRProblem}, the equilibrium of susceptible individuals $\bar s$ is always lower bounded by the equilibrium $\bar s^0$.}
\end{cor}
\emph{Proof} \EF{At equilibrium it must be that $\bar{i}=0$, and equation~\eqref{fSIRPeakNew} yields:
\[\log\frac{s_0}{\bar s}=\mathcal{R}_0(1 - \bar s) - \int_0^{\bar \tau} i \frac{\kappa(i)}{1+\kappa(i)}d\sigma, \] 
where $\bar \tau$ is the time it takes to reach equilibrium. The equilibrium $0\leq \bar s \leq 1$ must satisfy this equation. If $\kappa(i)=0$ for all $i$, we recover expression~\eqref{SIRss}: the left side of the equation is a curve that decreases monotonically as a function of $\bar s$, and the right side of the equation is a line with slope $-\mathcal{R}_0$ and intercept $\mathcal{R}_0$ when $\bar s=0$. If $\kappa(i)\not = 0$, the left side of the equation is unchanged. The right side is still a line with slope $-\mathcal{R}_0$, however it intercepts the $y-$axis at a point  $b<\mathcal{R}_0$, because the integral term is non-negative for any value of $\kappa$; this is equivalent to shifting down the line. Thus, when $\kappa(i)\not = 0$, the intersection point $\bar s$ of the curves on the left and right side of the equation intercept must be larger than the intersection when $\kappa(i)= 0$.  $\hfill \square$}

This proposition shows that, relative to an epidemic that lacks negative feedback, the fSIR model \EF{settles to a larger susceptible population in the disease-free equilibrium for any value of $\mathcal{R}_0$ and mitigation function}. As a consequence, the equilibrium recovered population satisfies $\bar r < \bar r^0$.

\EF{{\bf Linear mitigation function:}} In the case of mitigation function linearly proportional to infections, $\kappa(i)=\kappa\cdot i$, the fSIR model can be solved exactly in phase space \EF{as demonstrated in~\citep{capasso1978generalization} and~\citep{baker2020reactive}:}
\[\frac{di}{ds} =-1 + \frac{1+\kappa\cdot i}{\mathcal{R}_0s},\]
terms can be rearranged to find an ordinary differential equation for $i(s)$:
\[s\frac{di}{ds} -\frac{\kappa}{\mathcal{R}_0} i(s) = - s + \frac{1}{\mathcal{R}_0}.\]
With the change of variable $z=\ln(s)$, we find:
\[\frac{di(z)}{dz} -\frac{\kappa}{\mathcal{R}_0} i(z) = - \EF{e^{z}} + \frac{1}{\mathcal{R}_0},\]
which can be solved finding the phase-space expression:
\begin{equation} \label{ExactInfectionSol}
i(s) = \left(i_0+ \frac{1}{\kappa} + \frac{\mathcal{R}_0}{\mathcal{R}_0-\kappa} \right )s^{\frac{\kappa}{\mathcal{R}_0}} -  \frac{1}{\kappa} - \frac{\mathcal{R}_0}{\mathcal{R}_0-\kappa} s.
\end{equation}
In the particular case when $\kappa=\mathcal{R}_0$, the solution is $i(s)=(s-1)/\mathcal{R}_0 - s\ln s$. By setting $i(s)=0$ one can find the final size of the susceptible population. By substituting $i_{max}=\frac{\mathcal{R}_0 s_{crit}-1}{\kappa}$ in equation~\eqref{ExactInfectionSol}, one can derive $s_{crit}$:
\[ s_{crit}=\frac{1}{\mathcal{R}_0}\left(i_0\kappa(\mathcal{R}_0-\kappa) + 1 + \kappa - \frac{\kappa}{\mathcal{R}_0} \right)^{\frac{1}{1-\frac{\kappa}{\mathcal{R}_0}}}\]
and the corresponding infection peak can be found exactly; it can be verified that the infection peak always decreases with $\kappa$ as predicted by Proposition~\ref{PeakProp}.  

\EF{To the best of my knowledge, an exact solution of fSIR with linear mitigation function has not been found. However, similar models with other particular forms of the nonlinear transmission rate can be solved exactly~\citep{bohner2019exact}.}

I conjecture that in the presence of mitigation ($\kappa>0$) the time at which the infection peak occurs is always delayed (although moderately) relative to the SIR model. While this conjecture is corroborated by numerical computations, a formal proof is left for future work.


\subsection{Computational simulations}

In these computational simulations I consider the  fSIR model with linear mitigation function $\kappa(i)=\kappa\cdot i$ for illustrative purposes. It is assumed that $\kappa$ remains constant unless otherwise noted. 

\begin{figure}[htbp]
\centering
\includegraphics[width=0.7\columnwidth]{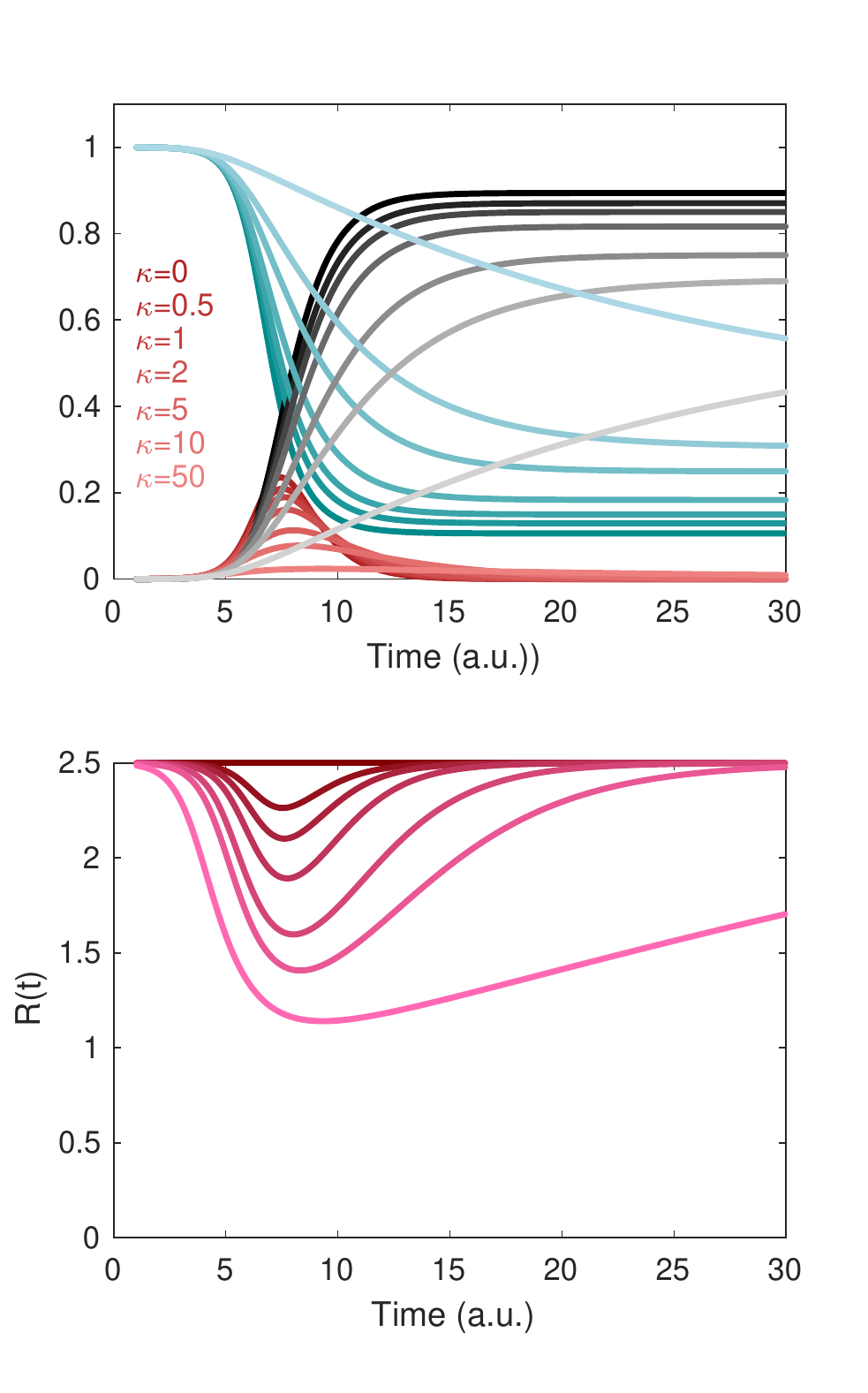}
\vspace{-.5cm}
\caption{Numerically integrated solutions of the fSIR model. Top: Susceptible (green), infected (red), and recovered (gray) individuals when the parameter $\kappa$ is varied (low to high, color shades from dark to light). Bottom: Evolution of the reproduction number in time computed from the simulations above; this can be interpreted as a qualitative measure of the implemented social distancing policies.}
\label{fSIRDynamics}
\end{figure}

Fig.~\ref{fSIRDynamics}, top, shows the numerically integrated solution of the  fSIR model~\eqref{fSIRn1}-\eqref{fSIRn2} with $\mathcal{R}_0=2.5$ as the parameter $\kappa$ is varied. ($\mathcal{R}_0=2.5$ corresponds to a choice of $\beta=0.25$ and  $\gamma =1/10$, i.e. the average time to recovery or death assumed to be 10 days; for comparison, the estimated average time to recovery in the COVID-19 epidemic is about 17 days for hospitalized patients~\citep{zhou2020clinical}). These simulations confirm that the peak of infections decreases with a large $\kappa$, relative to the case $\kappa=0$ (SIR without feedback). Fig.~\ref{fSIRDynamics}, bottom, shows the temporal evolution of the reproduction number in each simulation in the top panel: when infections increase, $\mathcal{R}(\tau)$ decreases; as infections decrease,  $\mathcal{R}(\tau)$ converges to the nominal level ($\mathcal{R}_0=2.5$).

\subsubsection*{The duration of an epidemic is extended \EF{in the presence of mitigation}}

Simulations in Fig.~\ref{fSIRDynamics} suggest that a large value of $\kappa$ extends the duration of the epidemic. This is evident by examining an approximation of the fSIR solution (Problem~\ref{fSIRProblem}): when $\kappa$ is very large, thus $\kappa\cdot i \gg 1$, the fSIR can be approximated by the linear system:
\begin{align}\label{fSIRnApprox}
\frac{d \hat s}{d\tau} & \approx -\frac{\mathcal{R}_0}{\kappa} \hat  s,\qquad\frac{d \hat i}{d\tau} \approx \frac{\mathcal{R}_0}{\kappa} \hat  s - \hat  i.
\end{align}
The solution $\hat  i (\tau)$ can be found exactly:
\begin{align}\label{fSIRnApproxSol} 
\hat  i (\tau) = i_0 e^{-\tau} + s_0\frac{\mathcal{R}_0}{\kappa-\mathcal{R}_0} \left(e^{-\tau} - e^{-\frac{\mathcal{R}_0}{\kappa} \tau} \right). 
\end{align}
This approximation shows that if ${\mathcal{R}_0}/{\kappa} \ll 1$ the infection dynamics converge very slowly  to $\hat i =0$ (convergence is dominated by the constant ${\mathcal{R}_0}/{\kappa}$).

\begin{figure}[htbp]
\centering
\includegraphics[width=0.8\columnwidth]{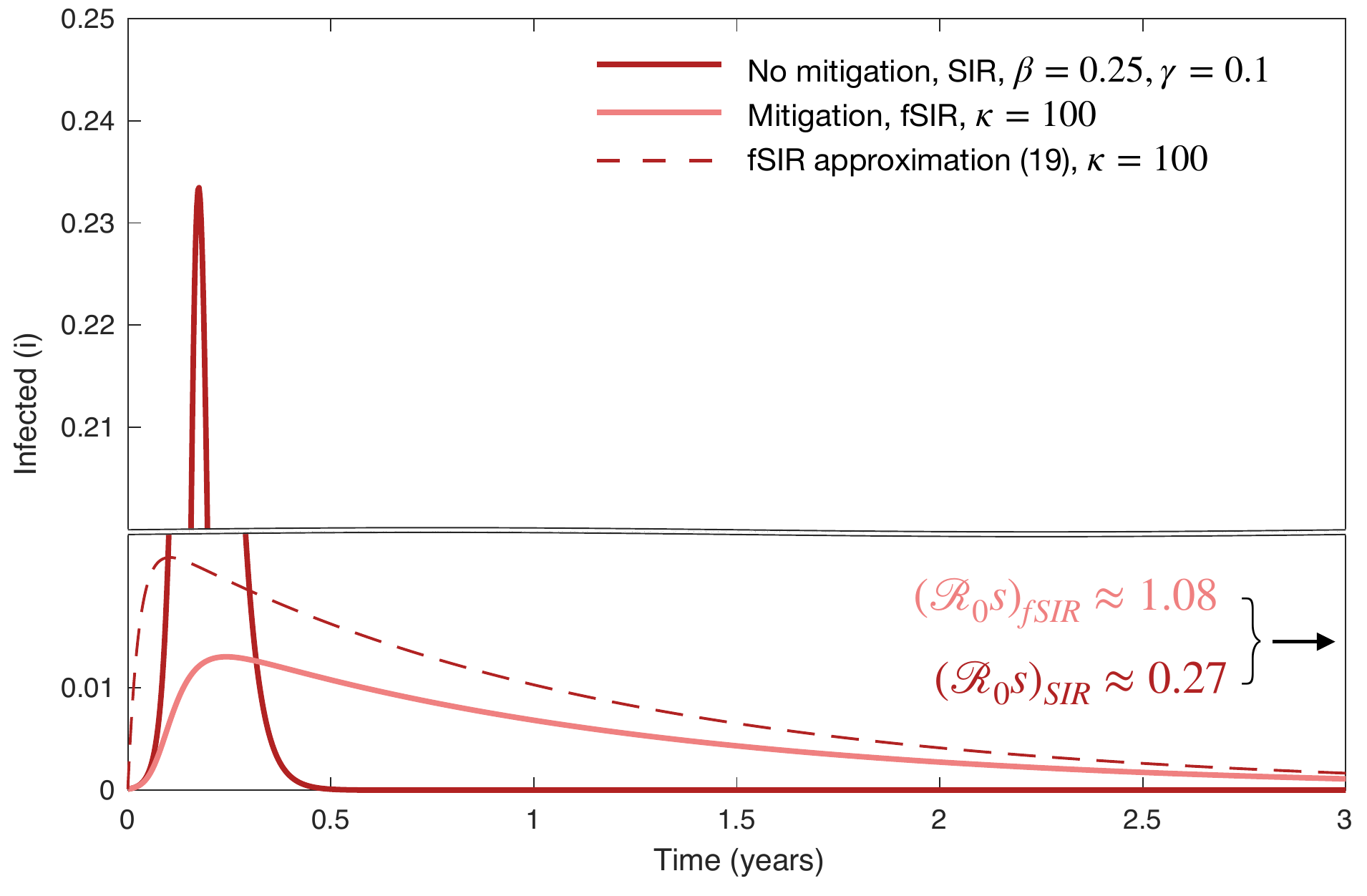}
\vspace{-.5cm}
\caption{Mitigation based on infection awareness extends the duration of an epidemic ($\beta=0.25$ and  $\gamma =0.1$). This simulation compares normalized infections in the SIR model with infections in the fSIR model ($\kappa=100$), and the fSIR linear approximation~\eqref{fSIRnApproxSol}. The y-axis is broken to emphasize the different timescale of convergence for SIR and fSIR. After 3 years, the SIR model does not admit a new outbreak; in contrast, if mitigation were to be completely relaxed ($\mathcal{R}_0=2.5$) the fSIR model could generate a new peak of infections because $\mathcal{R}_0 s >1$.} 
\label{fSIRConvergence}
\end{figure}

Simulations in Fig.~\ref{fSIRConvergence} compare infections in a SIR and fSIR model with focus on the timescale of convergence to the disease-free equilibrium. Cumulative infections under the unmitigated epidemic are higher than in the mitigated case. However, the unmitigated epidemic extinguishes in about 6 months; in contrast, the infection-aware mitigation strategy maintains a significant level of infectious individuals for a much longer time. Further, after 3 years, the unmitigated epidemic cannot generate another outbreak ($\mathcal{R}_0 s <1$), while the mitigated case may generate a new outbreak if social distancing and PPE were to be abandoned allowing $\mathcal{R}_0$ to return to its original value.    

\subsubsection*{Infection-aware mitigation strategies reduce the peak of infection and do not postpone the peak significantly}
\begin{figure}[htbp]
\centering
\includegraphics[width=0.7\columnwidth]{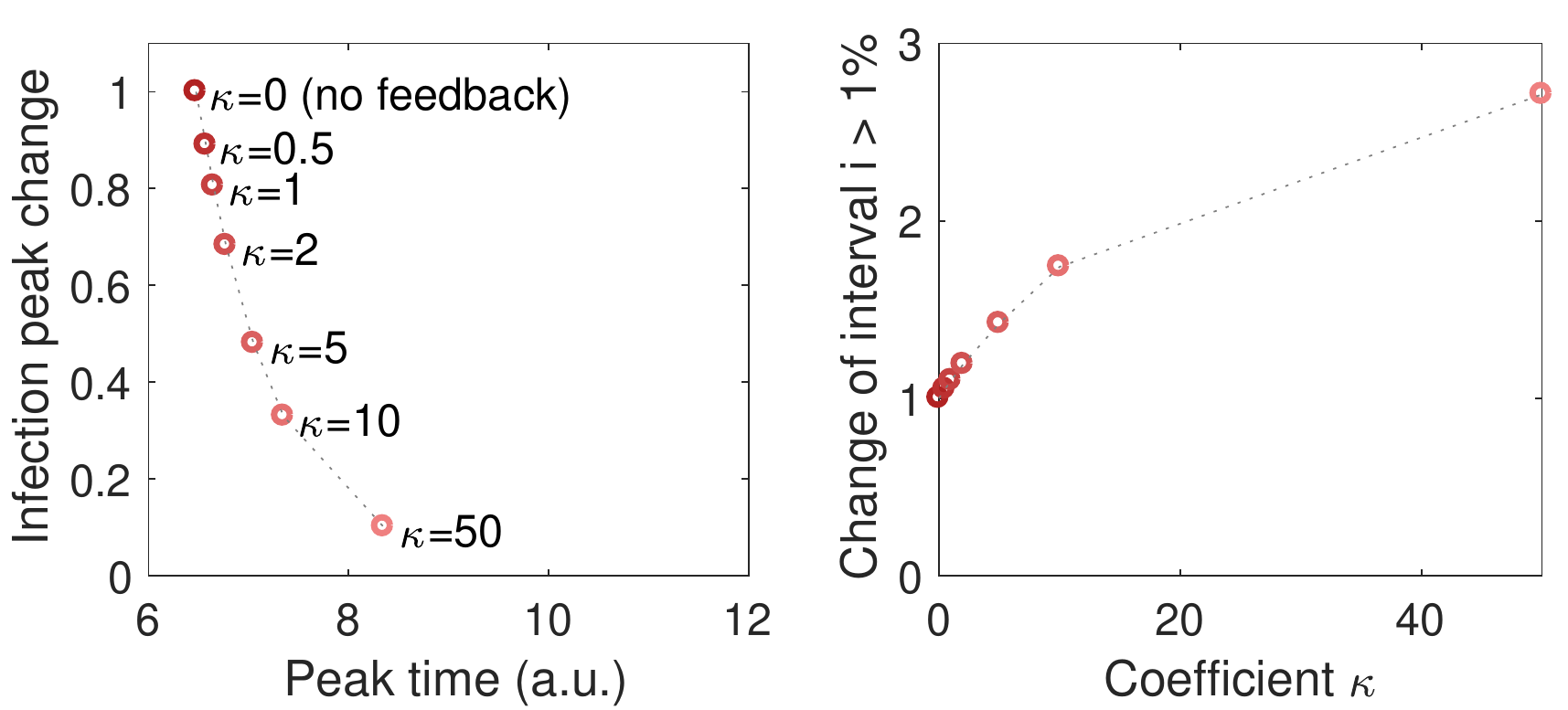}
\caption{Left: Peak time versus peak value of infections for different values of the feedback parameter $\kappa$. This plot evidences that the peak is not delayed as in models where the transmission rate is constant and low. Right: The duration of infections is longer in the presence of feedback; here it is measured as the time interval for which the fraction of infected individuals is larger than 2.5\% of the population.}
\label{PeakAnalysis}
\end{figure}

The simulations in Fig.~\ref{fSIRDynamics} confirm the results of Propositions~\ref{PeakProp}, because the infection peak is always reduced. Additional simulations in Fig.~\ref{PeakAnalysis} show that with a feedback parameter $\kappa=2$ (taken as an illustrative value) the infection peak size can be reduced by about 30\%, but this also causes a ~30\% extension of the time during which more than $2.5\%$ of the population is infected. This is consistent with the observation made earlier that the duration of the epidemic is extended when adopting infection-dependent mitigation policies.

\subsubsection{Effects of delayed infection awareness}

Delays in detecting and reporting infections are to be expected~\citep{li2020substantial}. \EF{While a theoretical analysis of the equilibria of the fSIR model with delays is not reported here, it may be pursued using local or global methods used for very similar models in~\citep{beretta1995global,huang2010global,kyrychko2005global,kumar2020deterministic,li2014sir}. Rather, computational simulations are used here to examine whether a delay $\Delta$ in obtaining infection information can compromise the effects of mitigation feedback.} A delay is included in the transmission rate expression: 
\EF{\begin{align}\label{fSIRnDelay}
\frac{ds}{d\tau}&= - \mathcal{R}(i(\tau-\Delta) ) s i, \qquad \mathcal{R}(i(\tau-\Delta))=\mathcal{R}_0\frac{1}{1+\kappa\cdot i(\tau-\Delta) },\\
\frac{di}{d\tau}&=(\mathcal{R}(i(\tau-\Delta) ) s -1) i. 
\end{align}}
While stability of this model with delay is not examined here, global stability analysis of SIR models with nonlinear transmission and delays have been demonstrated in~\citep{huang2010global}, and likely hold in this case. 

\begin{figure}[htbp]
\centering
\includegraphics[width=\columnwidth]{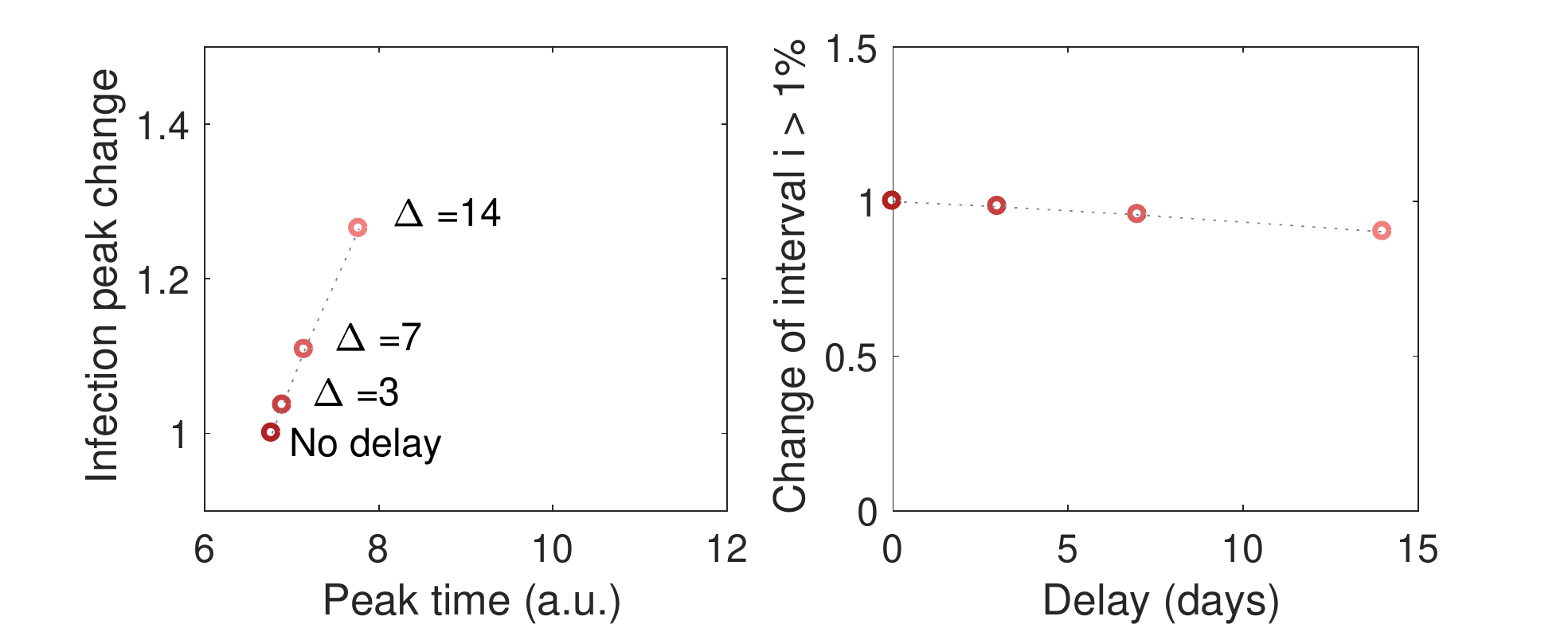}
\caption{Effects of delays on the peak size and duration. Left: Change in peak size in the presence of delays, relative to the case in which feedback is present without delay and $\kappa=2$. Right: The amount of time for which the fraction of infected population exceeds 2.5\% is slightly reduced when delays are between 0 and 14  days.}
\label{PeakAnalysisDelay}
\end{figure}

For illustrative purposes, I choose a feedback parameter $\kappa=2$ that remains fixed in these simulations, with $\mathcal{R}_0=2.5$ ($\beta=0.25$ and $\gamma=1/10$). Fig.~\ref{PeakAnalysisDelay} shows that a delay of up to 7 days  increases the peak by less than 10\%, but a 14 day delay causes a 25\% increase in the peak, offsetting the peak reduction obtained by introducing feedback (the simulated non-dimensional delay is divided by the rescaling constant $\gamma=1/10$). 

\subsubsection{The fSIR model captures the COVID-19 infection trends in the presence of mitigation strategies}\label{DataFitting}
The fSIR model was fitted to COVID-19 temporal series data for infections, recoveries, and deaths available from the Johns Hopkins Github repository~\citep{dong2020interactive}, last accessed on July 15, 2020. I selected data from four western democracies: Italy,  United Kingdom, Sweden, and the United States. The data were processed to compute active infections in a given day, and recoveries and deaths were summed and consolidated into the ``recovered'' compartment. All data were normalized by country population and thresholded to include only data collected after infections exceed 3 per million. Parameters were fitted with constraints $\beta\in[0\quad 0.6]$, $\beta\in[1/20 \quad 1/10]$, and $\kappa\in[0\quad 10\cdot 10^3]$; in the fitting score function, the infection prediction error was assigned a 100-fold penalty relative to the recovery data, with the expectation that recoveries may not be accurately reported for non-hospitalized patients. As a consequence, infection data are reproduced much more closely than recovery data by computationally generated trajectories that use fitted parameters. 

Initial epidemic data in Italy,  UK, Sweden, and the US are comparable, with reported infections and deaths showing similar doubling time of 2-4 days in the early (exponential) stages~\citep{bertozzi2020challenges}. Mitigation or suppression measures were not \emph{immediately} enacted, unlike countries such as South Korea, Japan, and Singapore that rapidly imposed lockdowns and contact tracing. (Timing and duration of initial interventions  are critical for a successful containment~\citep{sadeghi2020universal}.) 

\begin{figure*}[htbp]
\centering
\includegraphics[width=\textwidth]{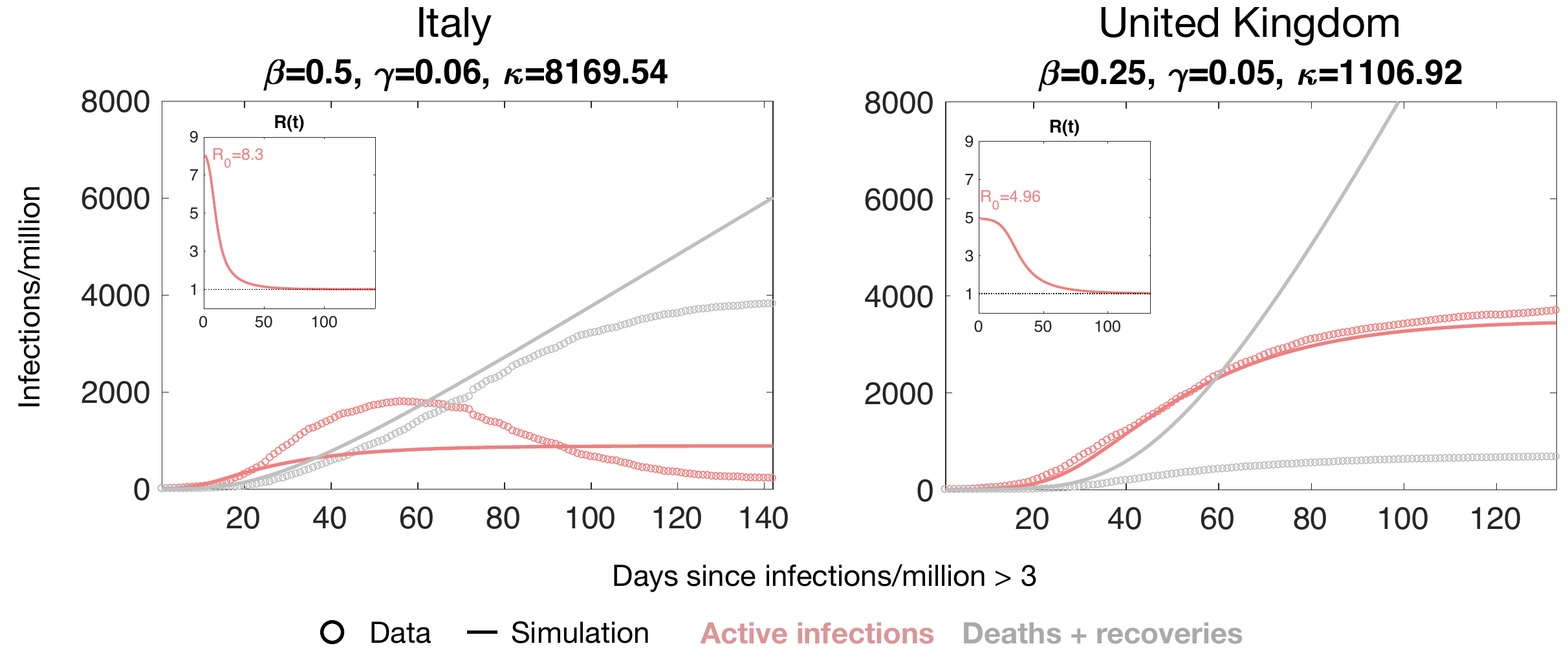}
\caption{COVID-19 active infection and recovery/death data for Italy and the United Kingdom, fitted using the fSIR model; the reproduction coefficient $\mathcal{R}(t)$ is shown in the insets. The fSIR model cannot capture the Italian infection data, as strict lockdown policies were enacted and enforced without relaxation for a sufficiently long time; this scenario would be better captured by a nearly discrete change in the reproduction coefficient (Fig.~\ref{SIRBeta}C). In contrast, the fSIR model reproduces very well active infection data in the UK, with realistic estimates for the transmission rate and $\mathcal{R}(t)$. This suggests that in practice, the UK strategy may be classified as an infection-based mitigation approach. Data fitting details are in Section~\ref{DataFitting}. }
\label{ItalyUK}
\end{figure*}

\begin{figure}[htbp]
\centering
\includegraphics[width=\columnwidth]{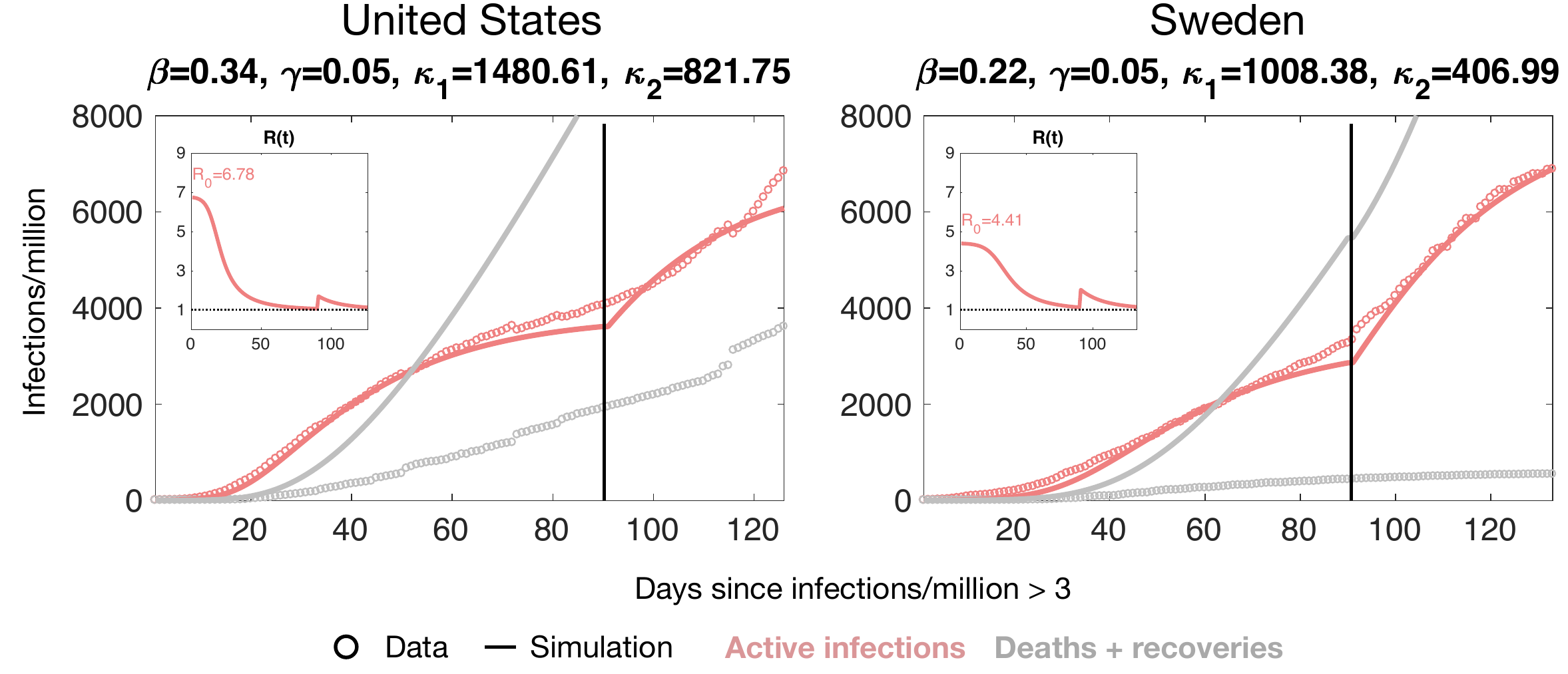}
\caption{COVID-19 active infection and recovery/death data for Sweden and the overall United States, fitted using the fSIR model, with reproduction coefficient $\mathcal{R}(t)$ in the insets. The infection trend of both countries shows two distinct phases, which are qualitatively separated by the black line placed at day 90. The fSIR model reproduces the trends observed adopting a different value of $\kappa$ in each phase ($\kappa_1$ for before day 90, $\kappa_2$ for after day 90). Data fitting details are in Section~\ref{DataFitting}. }
\label{SwedenUS}
\vspace{-.5cm}
\end{figure}

Italy is an example country that, like Spain and France, imposed and enforced a strict suppression strategy (lockdown), which resulted in a very limited number of new infections as of June 2020. While also the UK officially imposed lockdown/stay-at-home orders, their enforcement appears to have been less successful than Italy, as shown in Fig.~\ref{ItalyUK}.  From the beginning, Sweden  followed a mitigation strategy relying on personal responsibility of citizens to limit the spread of the virus, rather than on a strict lockdown strategy. Finally, the US is an example of a federal state in which disparate containment approaches were enacted at different times, from a tight lockdown in some states like New York and Michigan, to loose mitigation policies in other states like Arizona, Texas, and Florida. Interestingly, infection data from both Sweden and the US show a trend change around the end of May 2020, which is marked qualitatively by a black line at day 90 in Fig.~\ref{SwedenUS}. Because the overall US data includes contributions from all states, the first phase is likely dominated by the major outbreaks and lockdowns in the north eastern states in March and April 2020, while the second phase is dominated by southern states that relaxed mitigation strategies in May 2020. 

The fSIR model can cannot reproduce the infection data from Italy (in addition, the fitted transmission parameter $\beta$ is unrealistically high, and so is $\mathcal{R}_0$). Italy's COVID-19 reaction can be reproduced with a SIR model with a time-varying $\mathcal{R}_0$ tied to  fluctuations in lockdown measures~\citep{casella2020can}, that do not depend on infection levels (until new infections are nearly completely eliminated). In contrast, the fSIR model reproduces very well active infection trends in the UK, with realistic fitted parameters, suggesting that the UK lockdown measures were as effective as an infection-based mitigation strategy. The fitted value of $\kappa\approx1107$ means that a substantial societal reaction (reduction of the transmission coefficient) occurred relatively late in the epidemic, roughly when 0.1\% of the population was reported to be infected.

To fit infection data from the US and Sweden, we imposed single value of $\beta$ and $\gamma$ but allowed two distinct values of mitigation parameter $\kappa$ to capture the two apparent phases of the outbreak. In both cases, the fitted values of $\beta$ and $\gamma$ are realistic, and the values of $\kappa$ decrease in the second phase, suggesting that mitigation strategies were overall relaxed or that their effectiveness decreased over time. 

Even though all these countries ramped up their testing efforts, actual infection data are always underestimated. For this reason, it is interesting to test changes in the fSIR fitted parameters assuming a larger number of individuals affected by the epidemic. If data are scaled by X-fold (\emph{i.e.} infections and recoveries are believed to be X-times larger than reported), the fitted $\kappa$ qualitatively scales by a factor $1/X$, while changes in fitted $\beta$ and $\gamma$  are negligible.

This data fitting exercise has largely an illustrative purpose, and is not meant to put forward any predictions. The pitfalls of relaxing mitigation policies too early are discussed in detail using many models that are more complex and accurate than the one presented here~\citep{kissler2020projecting,giordano2020modelling}.

\section{Conclusion}
I have derived and examined the properties of a modified SIR model, here named feedback SIR (fSIR), in which infection-based mitigation policies introduce a reproduction number that decreases a continuous function of infection levels, generating a negative feedback loop. This simple model was originally described by~\citep{capasso1978generalization}, and here it is derived from first principles by considering cases in which individuals reduce their contacts or use PPE as more infections are reported. Using a time-scale separation argument, it was shown that the transmission rate function takes the form of a Holling type II or Michaelis-Menten function popular in ecology, chemistry and biology. It was demonstrated that mitigation based on infection awareness always reduces the infection peak, but substantially lengthens the duration of the epidemic. In the special case of a mitigation function that is linear with respect to infection information, this model requires only one additional parameter to capture the effects of social distancing and is amenable to exact analysis~\citep{baker2020reactive}. Extending the results presented here to an fSEIR model appears trivial, but is left for future work.
 
The reduction of transmission rate as a function of knowledge of infections, recoveries, or deaths goes beyond non-pharmacological mitigation strategies. While it is unlikely that vaccines for SARS-CoV-2 will be available before 2021, information about infection levels is likely to increase the likelihood of mass vaccination and thus cause a substantial decrease in the susceptible population; models like the one presented here may describe well this scenario~\citep{bootsma2007effect,kiss2010impact,buonomo2008global}. As widespread access to real-time epidemic information is available, and contact tracing becomes prevalent, closed-loop feedback regulation of epidemics is within reach. The role of nonlinear transmission parameters that introduce feedback is yet to be ascertained within more sophisticated compartment models developed for COVID-19 ~\citep{kissler2020projecting,giordano2020modelling}. While accurate forecasting will take advantage of complex models integrating data on multiple scales, simple models like the one presented here can provide general insights and guidelines to policymakers, doctors, and educators.

\section{Methods}
Differential equations were integrated with a forward Euler method in MATLAB using custom scripts, or using MATLAB's {\tt ode45}. Data fitting was done using MATLAB's {\tt fmincon}.

\section{References}

\bibliography{fSIRBibliography}

\begin{thebibliography}{47}
\providecommand{\natexlab}[1]{#1}
\providecommand{\url}[1]{\texttt{#1}}
\expandafter\ifx\csname urlstyle\endcsname\relax
  \providecommand{\doi}[1]{doi: #1}\else
  \providecommand{\doi}{doi: \begingroup \urlstyle{rm}\Url}\fi

\bibitem[Anastassopoulou et~al.(2020)Anastassopoulou, Russo, Tsakris, and
  Siettos]{anastassopoulou2020data}
C.~Anastassopoulou, L.~Russo, A.~Tsakris, and C.~Siettos.
\newblock Data-based analysis, modelling and forecasting of the {COVID-19}
  outbreak.
\newblock \emph{PloS one}, 15\penalty0 (3):\penalty0 e0230405, 2020.

\bibitem[Anderson and May(1978)]{anderson1978regulation}
R.~M. Anderson and R.~M. May.
\newblock Regulation and stability of host-parasite population interactions: I.
  regulatory processes.
\newblock \emph{The journal of animal ecology}, pages 219--247, 1978.

\bibitem[Baker(2020)]{baker2020reactive}
R.~Baker.
\newblock Reactive social distancing in a {SIR} model of epidemics such as
  {COVID-19}.
\newblock \emph{arXiv preprint arXiv:2003.08285}, 2020.

\bibitem[Beretta and Takeuchi(1995)]{beretta1995global}
E.~Beretta and Y.~Takeuchi.
\newblock Global stability of an {SIR} epidemic model with time delays.
\newblock \emph{Journal of mathematical biology}, 33\penalty0 (3):\penalty0
  250--260, 1995.

\bibitem[Bertozzi et~al.(2020)Bertozzi, Franco, Mohler, Short, and
  Sledge]{bertozzi2020challenges}
A.~L. Bertozzi, E.~Franco, G.~Mohler, M.~B. Short, and D.~Sledge.
\newblock The challenges of modeling and forecasting the spread of covid-19.
\newblock \emph{Proceedings of the National Academy of Sciences}, 2020.
\newblock \doi{10.1073/pnas.2006520117}.

\bibitem[Bin et~al.(2020)Bin, Cheung, Crisostomi, Ferraro, Myant, Parisini, and
  Shorten]{bin2020fast}
M.~Bin, P.~Cheung, E.~Crisostomi, P.~Ferraro, C.~Myant, T.~Parisini, and
  R.~Shorten.
\newblock On fast multi-shot epidemic interventions for post lock-down
  mitigation: Implications for simple {COVID-19} models.
\newblock \emph{arXiv preprint arXiv:2003.09930}, 2020.

\bibitem[Bohner et~al.(2019)Bohner, Streipert, and Torres]{bohner2019exact}
M.~Bohner, S.~Streipert, and D.~F. Torres.
\newblock Exact solution to a dynamic {SIR} model.
\newblock \emph{Nonlinear Analysis: Hybrid Systems}, 32:\penalty0 228--238,
  2019.

\bibitem[Bootsma and Ferguson(2007)]{bootsma2007effect}
M.~C. Bootsma and N.~M. Ferguson.
\newblock The effect of public health measures on the 1918 influenza pandemic
  in {US} cities.
\newblock \emph{Proceedings of the National Academy of Sciences}, 104\penalty0
  (18):\penalty0 7588--7593, 2007.

\bibitem[Buonomo et~al.(2008)Buonomo, {d'Onofrio}, and
  Lacitignola]{buonomo2008global}
B.~Buonomo, A.~{d'Onofrio}, and D.~Lacitignola.
\newblock Global stability of an {SIR} epidemic model with information
  dependent vaccination.
\newblock \emph{Mathematical biosciences}, 216\penalty0 (1):\penalty0 9--16,
  2008.

\bibitem[Calafiore et~al.(2020)Calafiore, Novara, and
  Possieri]{calafiore2020time}
G.~C. Calafiore, C.~Novara, and C.~Possieri.
\newblock A time-varying {SIRD} model for the {COVID-19} contagion in italy.
\newblock \emph{Annual reviews in control}, 2020.

\bibitem[Capasso and Serio(1978)]{capasso1978generalization}
V.~Capasso and G.~Serio.
\newblock A generalization of the {Kermack-McKendrick} deterministic epidemic
  model.
\newblock \emph{Mathematical Biosciences}, 42\penalty0 (1-2):\penalty0 43--61,
  1978.

\bibitem[Casella(2020)]{casella2020can}
F.~Casella.
\newblock Can the {COVID-19} epidemic be managed on the basis of daily data?
\newblock \emph{arXiv preprint arXiv:2003.06967}, 2020.

\bibitem[Chapwanya et~al.(2012)Chapwanya, Lubuma, and
  Mickens]{chapwanya2012enzyme}
M.~Chapwanya, J.~M.-S. Lubuma, and R.~E. Mickens.
\newblock From enzyme kinetics to epidemiological models with
  {Michaelis--Menten} contact rate: Design of nonstandard finite difference
  schemes.
\newblock \emph{Computers \& Mathematics with Applications}, 64\penalty0
  (3):\penalty0 201--213, 2012.

\bibitem[Dawes and Souza(2013)]{dawes2013derivation}
J.~Dawes and M.~Souza.
\newblock A derivation of {Holling's} type i, ii and iii functional responses
  in predator--prey systems.
\newblock \emph{Journal of theoretical biology}, 327:\penalty0 11--22, 2013.

\bibitem[Della~Rossa et~al.(2020)Della~Rossa, Salzano, Di~Meglio, De~Lellis,
  Coraggio, Calabrese, Guarino, Cardona-Rivera, De~Lellis, Liuzza,
  et~al.]{della2020network}
F.~Della~Rossa, D.~Salzano, A.~Di~Meglio, F.~De~Lellis, M.~Coraggio,
  C.~Calabrese, A.~Guarino, R.~Cardona-Rivera, P.~De~Lellis, D.~Liuzza, et~al.
\newblock A network model of italy shows that intermittent regional strategies
  can alleviate the {COVID-19} epidemic.
\newblock \emph{Nature communications}, 11\penalty0 (1):\penalty0 1--9, 2020.

\bibitem[Dong et~al.(2020{\natexlab{a}})Dong, Du, and Gardner]{Ensheng2020}
E.~Dong, H.~Du, and L.~Gardner.
\newblock An interactive web-based dashboard to track {COVID-19} in real time.
\newblock \emph{The Lancet}, 2020{\natexlab{a}}.
\newblock https://plague.com/.

\bibitem[Dong et~al.(2020{\natexlab{b}})Dong, Du, and
  Gardner]{dong2020interactive}
E.~Dong, H.~Du, and L.~Gardner.
\newblock An interactive web-based dashboard to track {COVID-19} in real time.
\newblock \emph{The Lancet infectious diseases}, 2020{\natexlab{b}}.

\bibitem[Funk et~al.(2009)Funk, Gilad, Watkins, and Jansen]{funk2009spread}
S.~Funk, E.~Gilad, C.~Watkins, and V.~A. Jansen.
\newblock The spread of awareness and its impact on epidemic outbreaks.
\newblock \emph{Proceedings of the National Academy of Sciences}, 106\penalty0
  (16):\penalty0 6872--6877, 2009.

\bibitem[Giordano et~al.(2020)Giordano, Blanchini, Bruno, Colaneri, Di~Filippo,
  Di~Matteo, and Colaneri]{giordano2020modelling}
G.~Giordano, F.~Blanchini, R.~Bruno, P.~Colaneri, A.~Di~Filippo, A.~Di~Matteo,
  and M.~Colaneri.
\newblock Modelling the {COVID-19} epidemic and implementation of
  population-wide interventions in italy.
\newblock \emph{Nature Medicine}, pages 1--6, 2020.

\bibitem[Greenhalgh et~al.(2015)Greenhalgh, Rana, Samanta, Sardar,
  Bhattacharya, and Chattopadhyay]{greenhalgh2015awareness}
D.~Greenhalgh, S.~Rana, S.~Samanta, T.~Sardar, S.~Bhattacharya, and
  J.~Chattopadhyay.
\newblock Awareness programs control infectious disease--multiple delay induced
  mathematical model.
\newblock \emph{Applied Mathematics and Computation}, 251:\penalty0 539--563,
  2015.

\bibitem[Harko et~al.(2014)Harko, Lobo, and Mak]{Harko14}
T.~Harko, F.~S.~N. Lobo, and M.~K. Mak.
\newblock Exact analytical solutions of the susceptible-infected-recovered
  ({SIR}) epidemic model and of the {SIR} model with equal death and birth
  rates.
\newblock \emph{Applied Mathematics and Computation}, 236:\penalty0 184--194,
  2014.

\bibitem[Hethcote(1976)]{hethcote1976qualitative}
H.~W. Hethcote.
\newblock Qualitative analyses of communicable disease models.
\newblock \emph{Mathematical Biosciences}, 28\penalty0 (3-4):\penalty0
  335--356, 1976.

\bibitem[Hethcote(2000)]{hethcote2000mathematics}
H.~W. Hethcote.
\newblock The mathematics of infectious diseases.
\newblock \emph{SIAM review}, 42\penalty0 (4):\penalty0 599--653, 2000.

\bibitem[Huang et~al.(2010)Huang, Takeuchi, Ma, and Wei]{huang2010global}
G.~Huang, Y.~Takeuchi, W.~Ma, and D.~Wei.
\newblock Global stability for delay {SIR} and {SEIR} epidemic models with
  nonlinear incidence rate.
\newblock \emph{Bulletin of mathematical biology}, 72\penalty0 (5):\penalty0
  1192--1207, 2010.

\bibitem[Kermack and McKendrick(1927)]{kermack1927contribution}
W.~O. Kermack and A.~G. McKendrick.
\newblock A contribution to the mathematical theory of epidemics.
\newblock \emph{Proceedings of the royal society of london. Series A,
  Containing papers of a mathematical and physical character}, 115\penalty0
  (772):\penalty0 700--721, 1927.

\bibitem[Kiss et~al.(2010)Kiss, Cassell, Recker, and Simon]{kiss2010impact}
I.~Z. Kiss, J.~Cassell, M.~Recker, and P.~L. Simon.
\newblock The impact of information transmission on epidemic outbreaks.
\newblock \emph{Mathematical biosciences}, 225\penalty0 (1):\penalty0 1--10,
  2010.

\bibitem[Kissler et~al.(2020)Kissler, Tedijanto, Goldstein, Grad, and
  Lipsitch]{kissler2020projecting}
S.~M. Kissler, C.~Tedijanto, E.~Goldstein, Y.~H. Grad, and M.~Lipsitch.
\newblock Projecting the transmission dynamics of {SARS-CoV-2} through the
  postpandemic period.
\newblock \emph{Science}, 2020.

\bibitem[Kolokolnikov and Iron(2020)]{kolokolnikov2020law}
T.~Kolokolnikov and D.~Iron.
\newblock Law of mass action and saturation in {SIR} model with application to
  coronavirus modelling.
\newblock \emph{Infectious Disease Modelling}, 2020.

\bibitem[Korobeinikov(2006)]{korobeinikov2006lyapunov}
A.~Korobeinikov.
\newblock Lyapunov functions and global stability for {SIR} and {SIRS}
  epidemiological models with non-linear transmission.
\newblock \emph{Bulletin of Mathematical biology}, 68\penalty0 (3):\penalty0
  615, 2006.

\bibitem[Korobeinikov and Maini(2005)]{korobeinikov2005non}
A.~Korobeinikov and P.~K. Maini.
\newblock Non-linear incidence and stability of infectious disease models.
\newblock \emph{Mathematical medicine and biology: a journal of the IMA},
  22\penalty0 (2):\penalty0 113--128, 2005.

\bibitem[Kruse and Strack(2020)]{kruse2020optimal}
T.~Kruse and P.~Strack.
\newblock Optimal control of an epidemic through social distancing.
\newblock 2020.

\bibitem[Kucharski et~al.(2020)Kucharski, Russell, Diamond, Liu, Edmunds, Funk,
  and Eggo]{earlyR0}
A.~J. Kucharski, T.~W. Russell, C.~Diamond, Y.~Liu, J.~Edmunds, S.~Funk, and
  R.~M. Eggo.
\newblock Early dynamics of transmission and control of covid-19: a
  mathematical modelling study.
\newblock \emph{The Lancet, Infectious Diseases}, 2020.
\newblock March 11, 2020.

\bibitem[Kumar et~al.(2020)Kumar, Goel, et~al.]{kumar2020deterministic}
A.~Kumar, K.~Goel, et~al.
\newblock A deterministic time-delayed {SIR} epidemic model: mathematical
  modeling and analysis.
\newblock \emph{Theory in Biosciences}, 139\penalty0 (1):\penalty0 67--76,
  2020.

\bibitem[Kyrychko and Blyuss(2005)]{kyrychko2005global}
Y.~N. Kyrychko and K.~B. Blyuss.
\newblock Global properties of a delayed {SIR} model with temporary immunity
  and nonlinear incidence rate.
\newblock \emph{Nonlinear analysis: real world applications}, 6\penalty0
  (3):\penalty0 495--507, 2005.

\bibitem[Li and Zhang(2017)]{li2017dynamic}
G.-H. Li and Y.-X. Zhang.
\newblock Dynamic behaviors of a modified {SIR} model in epidemic diseases
  using nonlinear incidence and recovery rates.
\newblock \emph{PLoS One}, 12\penalty0 (4):\penalty0 e0175789, 2017.

\bibitem[Li and Liu(2014)]{li2014sir}
M.~Li and X.~Liu.
\newblock An sir epidemic model with time delay and general nonlinear incidence
  rate.
\newblock In \emph{Abstract and Applied Analysis}, volume 2014. Hindawi, 2014.

\bibitem[Li et~al.(2020)Li, Pei, Chen, Song, Zhang, Yang, and
  Shaman]{li2020substantial}
R.~Li, S.~Pei, B.~Chen, Y.~Song, T.~Zhang, W.~Yang, and J.~Shaman.
\newblock Substantial undocumented infection facilitates the rapid
  dissemination of novel coronavirus {(SARS-CoV-2)}.
\newblock \emph{Science}, 368\penalty0 (6490):\penalty0 489--493, 2020.

\bibitem[Liu et~al.(1986)Liu, Levin, and Iwasa]{liu1986influence}
W.-m. Liu, S.~A. Levin, and Y.~Iwasa.
\newblock Influence of nonlinear incidence rates upon the behavior of {SIRS}
  epidemiological models.
\newblock \emph{Journal of mathematical biology}, 23\penalty0 (2):\penalty0
  187--204, 1986.

\bibitem[Liu et~al.(1987)Liu, Hethcote, and Levin]{liu1987dynamical}
W.-m. Liu, H.~W. Hethcote, and S.~A. Levin.
\newblock Dynamical behavior of epidemiological models with nonlinear incidence
  rates.
\newblock \emph{Journal of mathematical biology}, 25\penalty0 (4):\penalty0
  359--380, 1987.

\bibitem[Prasse et~al.(2020)Prasse, Achterberg, Ma, and Mieghem]{Prasse2020}
B.~Prasse, M.~A. Achterberg, L.~Ma, and P.~V. Mieghem.
\newblock Network-based prediction of the 2019-ncov epidemic outbreak in the
  chinese province hubei, 2020.
\newblock https://arxiv.org/pdf/2002.04482.pdf.

\bibitem[Rosini(2020)]{ItalyCOVID19}
U.~Rosini.
\newblock {COVID-19 Italia - Monitoraggio} situazione, 2020.
\newblock GitHub repository of data from Italy COVID-19 epidemic
  https://github.com/pcm-dpc/COVID-19.

\bibitem[Sadeghi et~al.(2020)Sadeghi, Greene, and Sontag]{sadeghi2020universal}
M.~Sadeghi, J.~Greene, and E.~Sontag.
\newblock Universal features of epidemic models under social distancing
  guidelines.
\newblock \emph{bioRxiv}, 2020.

\bibitem[Samanta and Chattopadhyay(2014)]{samanta2014effect}
S.~Samanta and J.~Chattopadhyay.
\newblock Effect of awareness program in disease outbreak--a slow--fast
  dynamics.
\newblock \emph{Applied Mathematics and Computation}, 237:\penalty0 98--109,
  2014.

\bibitem[Stewart et~al.(2020)Stewart, Heusden, and Dumont]{stewart2020control}
G.~Stewart, K.~Heusden, and G.~A. Dumont.
\newblock How control theory can help us control {COVID-19}.
\newblock \emph{IEEE Spectrum}, 57\penalty0 (6):\penalty0 22--29, 2020.

\bibitem[Weitz et~al.(2020)Weitz, Beckett, Coenen, Demory, Dominguez-Mirazo,
  Dushoff, Leung, Li, M{\u{a}}g{\u{a}}lie, Park, et~al.]{weitz2020modeling}
J.~S. Weitz, S.~J. Beckett, A.~R. Coenen, D.~Demory, M.~Dominguez-Mirazo,
  J.~Dushoff, C.-Y. Leung, G.~Li, A.~M{\u{a}}g{\u{a}}lie, S.~W. Park, et~al.
\newblock Modeling shield immunity to reduce {COVID-19} epidemic spread.
\newblock \emph{Nature medicine}, pages 1--6, 2020.

\bibitem[Yu et~al.(2017)Yu, Lin, Chiu, and He]{yu2017effects}
D.~Yu, Q.~Lin, A.~P. Chiu, and D.~He.
\newblock Effects of reactive social distancing on the 1918 influenza pandemic.
\newblock \emph{PloS one}, 12\penalty0 (7), 2017.

\bibitem[Zhou et~al.(2020)Zhou, Yu, Du, Fan, Liu, Liu, Xiang, Wang, Song, Gu,
  et~al.]{zhou2020clinical}
F.~Zhou, T.~Yu, R.~Du, G.~Fan, Y.~Liu, Z.~Liu, J.~Xiang, Y.~Wang, B.~Song,
  X.~Gu, et~al.
\newblock Clinical course and risk factors for mortality of adult inpatients
  with covid-19 in wuhan, china: a retrospective cohort study.
\newblock \emph{The Lancet}, 2020.

\end{thebibliography}

\end{document}